# Compressed Sensing for Sparse Underwater Channel Estimation: Some Practical Considerations

Sushil Subramanian
*University of Southern California, Los Angeles, California*

**Abstract**

We examine the use of a structured thresholding algorithm for sparse underwater channel estimation using compressed sensing. This method shows some improvements over standard algorithms for sparse channel estimation such as matching pursuit, iterative detection and least squares.

## 1. Haupt and Nowak's Algorithm

In Ref. [1], Haupt and Nowak propose a method to recover signals corrupted with noisy random projections. Ref. [1] is an improvement over Ref. [2] where Candes and Tao propose a method called the Dantzig Selector to recover signals using random projections with a measurement matrix of lesser rank than the input signal's basis. In the Dantzig Selector problem, however, the signals are corrupted by bounded perturbations; Ref. [1] extends this to unbounded Additive White Gaussian Noise (AWGN) which is more practical in nature.

Consider a signal $x$ that can represented in an orthonormal basis $\Psi$. If $\Psi$ is a matrix and $x$ is considered to be a signal vector, they can be related as:

$$x = \Psi \theta \qquad (1)$$

If the vector $\theta$ has very few non-zero taps, the signal $x$ is said to be sparse in the basis $\Psi$. From work in Ref. [3] and [4], such a signal can be recovered by measuring it with an orthonormal matrix of much lesser rank than the basis itself (say the basis has rank $N$). This method is known as *compressed sensing*. In essence, if we consider a random orthonormal matrix $\Phi$ of size $K \times N$ with $K \ll N$, we may measure the signal $x$, at rate much lesser than that of the Shannon Rate, to obtain $y$ as:

$$y = \Phi x = \Phi \Psi \theta = A\theta \qquad (2)$$

As described in Ref. [3] and [4], to successfully recover the signal, the effective measurement matrix $A$ should satisfy the *restricted isometry condition* which puts some restriction on the orthogonality of $A$. Also, if the matrices $\Psi$ and $\Phi$ are largely *incoherent* the recovery probability is larger. Further, with large probability, the signals can be recovered by standard methods such as Basis Pursuit from l1 norm minimization [4] using techniques like Matching Pursuit [5], Simplex and Primal-Dual Interior Point Methods [6].

In Ref. [1], the nature the signal $x$ is allowed to be corrupted with Gaussian noise. Consider a signal measured as:

$$y = \Phi(x + \eta) + w \qquad (3)$$

where $w$ and $\eta$ are vectors consisting of random AWGN samples. The length of $w$ is $K$ and the length of $\eta$ is $N$. With this assumption, the l1 norm minimization mentioned before for the Dantzig Selector can be adapted to an l2 norm minimization to recover the signal, with a *risk of candidature* $\hat{R}(x)$ [1]:

$$\hat{x}_k = \arg\min_x \left\{ \hat{R}(x) + \frac{c(x)\log 2}{k\varepsilon} \right\} \tag{4}$$

where:

$$\varepsilon = \frac{1}{50(B+\sigma)^2}$$

$c(x) = 2S\log(N)$

$B$ = maximum possible signal power

$\sigma$ = noise variance

$S$ = Sparsity of the signal (number of non-zero taps in $\theta$)

$\hat{R}(x)$ = Risk of candidature of the received signal, an l2 norm function of the error

This problem can be simplified for the estimate of the sparse vector ($\hat{\theta}_k$) instead with the following modification:

$$\hat{\theta}_k = \arg\min_\theta \left\{ \|y - \Phi\Psi\theta\|^2 + \frac{2\log(2)\log(N)}{\varepsilon}\|\theta\|_0 \right\} \tag{5}$$

To solve the above system of equations, the following algorithm is used [1]:

$$\chi^{(t)} = \theta^{(t)} + \frac{1}{\lambda}(\Phi\Psi)^T \left( y - \Phi\Psi\theta^{(t)} \right)$$

$$\hat{\theta}_k^{(t+1)} = \begin{cases} \chi_k^{(t)} & \text{if } |\chi_i^{(t)}| \geq G = \sqrt{\frac{2\log(2)\log(N)}{\lambda\varepsilon}} \\ 0 & \text{otherwise} \end{cases} \tag{6}$$

where $\lambda$ is the largest eigenvalue of the matrix $\Phi$.

Note that Eq. (10) follows a two step process with an estimation of the possible vector $\chi$ and a soft decision step where the value of the threshold, $G$ is dependent highly on $\sigma$ and therefore the noise level. Here, $t$ determines the iteration number and $k$ is the tap number ($k$ = 1, 2, 3…., $N$).

The Dantzig Selector (DS) considers random perturbations. Therefore, implementations of the algorithm do not give good results for recovery of signals corrupted with AWGN resulting in SNR lesser than 40 *dB* [7]. However, Haupt and Nowak's Algorithm (HN) recovers signal considerably well till up to 20 *dB*, after which the signal are not recoverable. Also, since the algorithms are statistical, there is always a negligible probability Dantzig Selector and HN fail. Besides, HN is performed in $O(N^2K)$; however Dantzig Selector in $O(N^3)$.

## 2. Sparse Channel Estimation using Compressive Sampling

With many algorithms available for compressive sampling, it is clear that Haupt and Nowak's Algorithm is well suited for estimation problems in wireless communications as the paper considers AWGN. In Ref. [8], Bajwa et al. consider the application of Compressive Sampling to Sparse Channel Estimation. However, Ref. [8] does not discuss in detail the metrics used to evaluate the estimate of the sparse channel using their methods. Common metrics for the evaluation of the channel estimate are *Mean Square Error (MSE)* and *Bit Error Rate (BER)* of an equalizer output. Also, Ref. [8] considers upto 10 taps in a channel response of delay spread 128, which is not always the case with sparse channel. We will now describe qualitatively the channel model and look at previous work in Sparse Channel Estimation.

### A. *Sparse Underwater Acoustic Channels*

In practical situations, underwater acoustic (UWA) and ultra-wideband (UWB) channels exhibit a highly sparse channel response (in the time domain). Research in Ref. [17] shows that a typical channel response has very few non-zero taps but has a large delay spread. A typical channel shown in Fig. 1 has a delay spread of 100 taps with as few as 3 non-zero taps.

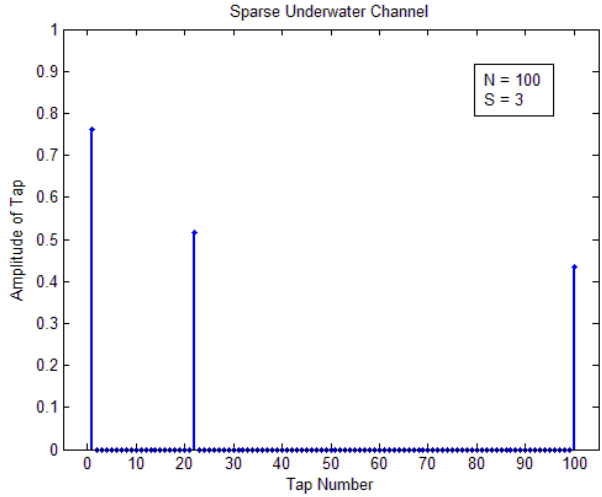

Figure 1. Time Domain response of a typical sparse underwater channel. Note the number of non-zero taps is as less as 3 in a delay spread of 100.

The problem of sparse channel estimation has been addressed in many previous works ([9]-[11]). References [9] and [10], attempt to solve the problem using the tested techniques of Matching Pursuit and Orthogonal Matching Pursuit (a variant of Matching Pursuit). Ref. [11] follows a different approach where many parallel structured estimates are performed after taking hard decisions on the channel tap locations. Such an estimate is termed *structured*, as a hard estimate determines first the possible tap locations and then uses pseudo-inverses of this matrix to estimate the amplitude at these tap locations. Accordingly, the method proposed in [11] performs significantly better than those in [9] and [10]. However, the number of computations is increased as there are many parallel estimation and decision loops based on noise information from an initial unstructured estimate (for details, see Ref. [11]).

Unlike compressive sampling, these methods use the entire received signal and not a projection of lesser rank to estimate the sparse channel. The property of reducing the rank using compressive sampling can thus serve as an advantage in reducing the amount of computation. Since compressive sampling techniques require random projection on a signal (see Ref. [4]), the methods will also serve to reduce the constraints of the measurement matrix. The random projections are usually random matrices of Gaussian or Rademacher Distributions, which satisfy the Restricted Isometry and Incoherence conditions (with nearly all possible basis expansion models [12]) as mentioned in Section 1. We may thus summarize the motivation for using compressive sampling in sparse channel estimation with the following points:

- Sparse channels can be estimated using compressive sampling if the algorithm used by the estimator satisfies the restrictions posed by the underlying theory

- Unlike techniques as used in Ref. [9]-[11], compressive sampling requires a project signal with a rank much lesser than that of the received signal (i.e. $K \ll N$) thus reducing the number of sensors and the computation complexity.

- Since the projection of the received signal is random, the measurement matrix need not be constrained by the input signal and noise power.

### B. Brute Use of Compressive Sampling

We will now consider the use of DS and HN for sparse channel estimation. Consider a system where a training sequence $c$ of length $M$ is transmitted through a channel with response $h$ and added AWGN $n$. The received signal at the receiver $r$ can be written as:

$$r = (C * h) + n \qquad (7)$$

The matrix $C$ is the convolution matrix formed from the training sequence. $C$ is a full Toeplitz matrix with dimensions $(M+N-1) \times (N)$ of the form:

$$C_{(M+N-1)\times(N)} = \begin{pmatrix} c(1) & 0 & \ldots & 0 \\ c(2) & c(1) & \ldots & 0 \\ \vdots & \vdots & \ddots & \vdots \\ c(M) & c(M-1) & \ldots & c(M-N+1) \\ 0 & c(M) & \ldots & c(M-N+2) \\ 0 & 0 & \ldots & c(M-N+3) \\ \vdots & \vdots & \ddots & \vdots \\ 0 & 0 & \ldots & c(M) \end{pmatrix} \quad (8)$$

The usual choice for the elements of the training sequence is a Rademacher distributed sequence of ±1 BPSK symbols. We will consider such a sequence for $c$. Since $h$ is sparse in nature, it makes sense to modify $C$ in order for it form an appropriate basis for the estimation of $h$. Note that with $C$ as basis, (7) approaches a form equivalent to (3), with $\eta = n$, $w = 0$, $x = \Psi\theta = Ch$. The received signal can be measured with a matrix $\Phi$ and the resulting vector $y$ can be now used to estimate $h$. Note that in the above model, to normalize the basis, we have assumed that the matrix $C$ is full Toeplitz, and there exists a guard interval of length $(N-1)$ before further data symbols are transmitted.

Note that to use a $C$, it must be orthonormal. Also, $\Phi C$ should satisfy the restricted isometry condition. Ref. [8] shows that with certain conditions on the parameters involved in DS, the matrix $A = \Phi C$ satisfies all properties required for use with compressive sampling methods. Similar proofs apply for other compressive sampling techniques, such as Lasso [13] and HN. Since $C$ is Rademacher distributed, the norm of the vectors is $\sqrt{\sum_{i=1}^{M}(\pm 1)^2} = \sqrt{M}$. If vectors in $C$ are normalized then $C$ becomes orthonormal. Thus using $C/\sqrt{M}$ instead of $C$ satisfies all conditions for the use of recovery techniques like HN, DS, Lasso etc.

We now apply DS and HN for the estimation of $h$, with the equivalent signal model as described above. The DS is l1 minimization technique [2] of the form:

$$\arg\min \|\theta\|_{l_1} \quad \text{subject to} \quad \|A^T(A\theta - y)\|_{\infty} \leq \gamma \quad (9)$$

which can be solved using the Primal-Dual Interior Point Method ([6], [7]).

Some changes are made to the HN algorithm. Ref. [1] suggests that the value of $G$ fluctuates highly, as it is related directly to $\sigma$ (noise variance), $N$ (delay spread) and $M$ (training sequence length). Also, in the proof of the use of the algorithm shown in (6), there is a constraint on the maximum permissible power of the signal $x$. This condition has to be satisfied in order for the algorithm in (6) to be used. We will consider the case of the threshold and the power constraint in the following section.

C. *Power Restrictions in Application of HN*

Consider the signal $x$ uncorrupted with AWGN. In the case of the convolution matrix, $x = Ch$. If $C$ is of the form as shown in (8), $x$ is a vector of length $(M+N-1)\times(N)$. The power restriction to use (6) as per [1] for the case of the signal model described above is:

$$\sum_{i=1}^{M+N-1}(x_i)^2 \equiv \|x_i^*\|^2 \leq (M+N-1)B^2 \quad (10)$$

Each element of the vector $x$ can be written as:

$$x_i = \sum_{j=1}^{N} C(i,j) * h(j) \quad i = 1,2,3,\ldots,(M+N-1) \quad (11)$$

$C$ is a Toeplitz matrix with consisting of random numbers from the Rademacher distribution. The mean of the distribution is 0 and the variance is 1. Assuming $h$ is a uniformly distributed or Gaussian (most channels responses are combinations of these two distributions), the mean of *h* is 0 and the variance is a non-zero value depending on the probability distribution function of *h* and the number of divisions between 0 and 1 in the time axis.

Suppose the sparsity of *h* is equal to *S*, the sparse representation of the individual $x_i$ will only involve a summation over *S* multiplications of *C(i,j)* and *h(j)*. With this definition of *x* we have to prove the inequality shown in (10). We will use the Hoeffding's Inequality ([14], [15]), which states that *k* independent bounded random variables $z_j$ with the condition $a_j \leq z_j \leq b_j$ satisfy:

$$\Pr\left((s - \mathrm{E}(s)) \geq t\right) \leq \exp\left(\frac{-2t^2}{\sum_{j=1}^{N}(b_j - a_j)^2}\right) \quad (12)$$

where $s = \sum_{j=1}^{N} z_j$. In the case of Eq. (11), substitute $z_j = C(i,j) * h(j)$. We see that $s_i = x_i = \sum_{j=1}^{N} C(i,j) * h(j)$. The variables *h(j)* and *C(i,j)* are bounded between 0 and 1, and between 0 and $1/\sqrt{M}$ respectively. This implies that $\left(0 \leq C(i,j) * h(j) \leq \left(1/\sqrt{M}\right)\right)$ which are the bounds $b_j$ and $a_j$. Since the mean of both elements of *C* and *h* is 0 and the variables are independent, $\mathrm{E}(s) = 0$. Since the values of power are all normalized to 1, we are interested in $t = \sqrt{1/M}$. Therefore, for the case of the signal model in consideration, it is easy to verify that:

$$\Pr\left(|x_i|^2 \geq 1/M\right) \leq \exp\left(\frac{-1}{2S}\right) \quad (13)$$

With the normalization factor introduced, we obtain a relation equivalent to both sides of Eq. (13) multiplied by *(M)* which results in the same inequality. The norm value of vector *x* is defined as in Eq. (10). Since the condition after normalizing is $x_1^2 + x_2^2 + ... + x_{M+N-1}^2 \leq \left(\frac{M+N-1}{M}\right)$, it is sufficient to prove the probability condition (13). With a simple convolution of identical probability distribution functions we show that:

$$\Pr\left(\sum_{i=1}^{M+N-1} |x_i|^2 \leq \left(\frac{M+N-1}{M}\right)\right) \leq 1 - \left[\exp\left(\frac{-1}{2S}\right)\right]^{M+N-1} \quad (14)$$

According to Ref. [15], the sparsity should satisfy $S \leq c\left(\sqrt{(M+N-1)/\log N}\right)$ where *c* is a constant. If we consider a fixed low sparsity *S*, as is typical in UWA channels described above, the probability in (14) can be made very high thus statistically ensuring that with most of $x_i$, the power constraint posed by Ref. [1] is satisfied. Thus, with very low sparsity and large delay spreads (as is with UWA channels), the compressive sampling technique described in [1] can be adapted to the estimation of highly sparse UWA channels. To show that the power constraints are satisfied, we consider a variable training sequence length, a delay spread of 100 and a sparsity of 3. It was found that on average, 86.3% of the experiments (out of 1000 experiments) satisfied the power constraint. Thus, the HN algorithm has chances of failure in some cases. This failure rate can be manipulated, if the training sequence is designed, in which case the probability limit in (14) will not hold. We will discuss failure of estimator in more detail later in this report.

*D. Experiments*

We now compare results of DS and HN with respect to the structured and unstructured Cramer-Rao Bounds (CRB). The unstructured and structured CRBs are defined as:

$$\text{CRB-U} = \left(\sigma^2\right) \text{trace}\left\{\left(C^T C\right)^{-1}\right\} \quad (15)$$

$$\text{CRB-S} = (\sigma^2)\,\text{trace}\left\{\left((Cdiag(\hat{b}))^T (Cdiag(\hat{b}))\right)^\dagger\right\} \qquad (16)$$

where † denotes the pseudo-inverse and $\hat{b}$ is a hard estimate of the time domain channel response. It is clear that CRB-S assumes a prior knowledge of the tap positions.

Note from Eq. (6), the value of $G$ is dependent on SNR through $\varepsilon$. Thus there is some uncertainty as to what would be the best threshold to work well for all SNRs. In Ref. [1], Haupt and Nowak made approximations and determined a lower bound on $\varepsilon$. After determining this value they reported that $G/4.6$ as the best possible threshold. In Ref. [1], the matrices were assumed ideal and the basis are deterministic. In the case of a Random Toeplitz Matrix as a basis ($C$), it is difficult to determine the threshold theoretically. Therefore, various values of a divisor $P$ were experimented with at all SNRs to obtain the best estimate using the threshold $G/P$. Fig. 2(a) shows the variation of the value of $P$ with SNR as observed empirically. Fig. 2(b) shows the MSE performance of DS and HN with respect to CRB-U and SRB-S for sparse channel estimation.

For the experiment, we have considered a training sequence of length, $M = 200$, with a delay spread of $N = 100$, and sparsity $S = 3$. In the DS algorithm, we have used Eq. (9) with $\gamma$ equal to 0.24, as suggested by analysis in Ref. [8]. In both the algorithms, the noise is AWGN. In HS, the measurement matrix $\Phi$ is a random Rademacher distributed matrix of dimensions $50 \times (M+N-1) = 50 \times 299$.

Observe that the plot of $P$ (or the optimized threshold ratio value) is random with respect to the SNR. Therefore, without prior knowledge of SNR it is quite impossible to determine the optimal threshold to use for sparse channel estimation. Also, observe that even though the decision step in (6) considers the SNR as a factor in the threshold, it is not tuned to the channel response as is the case of accurate estimators such as in [9] and [11]. However, taking optimal $P$ (as per Fig. 2(a)) and estimating using HN produces results closer to CRB-S than DS. This is due to the presence of the decision step in (6) that aims to provide some tradeoffs between bias and variance of the estimator. Also notice in HN that the power of the signal being an integral constraint of the algorithm tends to fluctuate more if the SNR is as high as 0 *dB*; thus poorer estimates at low SNR.

### 3. Structured Thresholding for HN

*A. Analysis of the Threshold in HN for Sparse Underwater Channels*

A analysis of the optimum threshold, with adaptation to the signal model discussed in Section 2 and considering a sparse underwater channel is now presented. The decision step threshold in (6) is given by:

$$G = \sqrt{\frac{2\log(2)\log(N)}{\lambda \varepsilon}} \qquad (17)$$

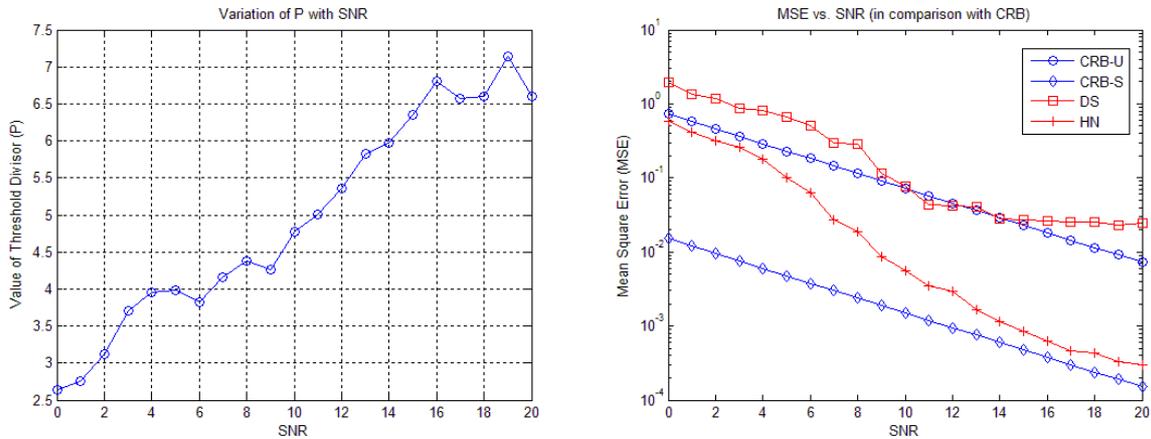

Figure 2. (a) The variation of the divisor for the threshold $P$ with SNR. Notice the increasing trend, but this threshold is an empirical estimate and is not the same for each experiment. (b) The MSE of HN and DS with respect to CRBs.

where $\lambda$ is the largest eigenvalue of the measurement matrix $\Phi$. The other parameters are as described in Eq. (4). Note that $\Phi$ has dimensions $K \times (M+N-1)$, and is a normalized Rademacher distributed random matrix. For $N = 100$ and $M = 200$, we obtain an appropriate maximum allowable sparsity $S = 8$ (from discussion in Section 2, Subsection C). We consider a much lesser sparsity, $S = 3$. For this sparsity it suffices to have a measurement rank [4]:

$$K \geq S \log(N) \tag{18}$$

In most channel estimation systems such as in [9] and [11], the delay spread is known. Therefore, assuming that the designer has knowledge of $N$, consider $K = N/2$. For $S = 3$, the relation (18) is satisfied for $N \geq 50$, thus covering all possible large delay spreads. The variation of $\lambda$ of a random Rademacher matrix of size $(N/2) \times (M+N-1)$ (and normalized by dividing by $\sqrt{N/2}$), with $N$ is shown in Fig. 3, where $M = 200$.

According to experiments and Ref. [16] it can be shown that this variation is independent of the probability distribution function of the matrix, thus compatible with the theory of compressive sampling where any orthonormal random matrix may be used to measure the signal. Also Ref. [16] suggests that the eigenvalue is a normal function with sharp falls around the mean, thus implying that the variation is not large with randomness of the matrix. Therefore, the plot in Fig.(3) is *more or less universal* for a given value of $M$. Assuming a region of interest $50 \leq N \leq 150$, the plot in Fig. (3) may be approximated to a linear approximation of the form:

$$\lambda(N) = 18.81 - 0.064(N) \tag{19}$$

It is clear that better approximations of the plot in Fig. (3) may be obtained with higher degree polynomials.

However, for simplicity we consider a linear approximation. Also note that the nature of the plot in Fig. (3) is not different with varying $M$, however different constants and slopes are obtained in Eq. (19). Substituting values of $\lambda$ and $\varepsilon$, the threshold $G$ may now be written as:

$$G = \left( \frac{\left(\sqrt{2\log(2) \times 50}\right)\sqrt{\log(N)}}{18.81 - 0.064N} \right)\left( \sigma + \frac{1}{\sqrt{M}} \right) \tag{20}$$

Substituting values $M = 200$ and $N = 100$, we obtain $G = 1.4397(\sigma + 0.0707)$. This expression of G is linear with respect to the variance in noise $\sigma$, and can be manipulated in the estimation process. In general, with linear approximations:

$$G = (a + \sigma)b \quad a,b \text{ are } f(M, N) \tag{21}$$

the constants $a$ and $b$ can be determined from the method described above.

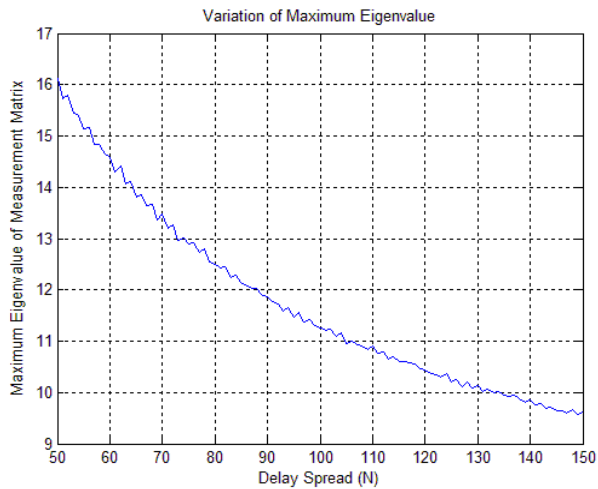

Figure 3. Variation of $\lambda$ with the delay spread $N$ for a measurement rank $K$.

## B. The initial channel estimate and the threshold set

A strong point of the algorithm presented in [11] is that it uses thresholds based on an initial estimate of the channel using a sliding correlator. This initial estimate can be represented as:

$$\hat{h}(n) = \frac{c_n^T r}{\|c_n\|^2}, \qquad n = 1, 2, ..., N \qquad (22)$$

where $c_n$ is each column of the convolution matrix $C$.

However, in the compressive sampling situation, the initial estimate used is generally a maximum energy estimate defined as $\hat{h} = A^T y$ [7], where $A = \Phi C$ and $y$ is the received signal after random projection. We may also define a sliding correlator similar to (22), with the columns of $A$ instead of $C$, and the signal $y$ instead of $r$. For the sparse channel response shown in Fig. 1, the sliding correlator and the maximum energy plots are shown in Fig. 4.

The plots show that both the estimates nearly match, as the matrices are all normalized. However, due to the random measurement matrix, both the estimates are highly corrupted with noise, in comparison non compressive sampling cases. With the threshold expression as defined by Eq. (20) and Eq. (21) and the initial channel estimate we aim to arrive at an optimal set of thresholds {**G**}, which can be used by the algorithm described in (6), instead of the arbitrarily taking a threshold value as defined by Fig. 2(a). By developing this method we also aim to find an optimal threshold set, which combines the use of the initial estimate as shown in [11] and the expression developed in Eq. (20) and (21). This will ensure that the estimation is tuned to both SNR and channel response.

An algorithm is defined as follows:

- Compute the initial estimate $\hat{h}^{(0)}$ as defined by (22).
- Consider the SNR of interest range to be $L \leq (SNR) \leq H$; the corresponding $\sigma$ values are $\sigma_l \leq \sigma \leq \sigma_h$.
- Using the empirical method defined in Subsection A, compute the function which gives $G = f(M,N,\sigma)$.
- For a known value of $N$, $M$ and $\sigma$, compute the function $G$ in the range $[\sigma_l, \sigma_h]$.
- Make the function $G$ discrete by quantizing $G$ to obtain $\{G_1, G_2, ..., G_{N_t}\}$.
- Similarly divide $\hat{h}$ into equal divisions to obtain $\Delta = \max_n \{|h(n)|\}$.
- For each $i$ going from 1 to $N_t$ calculate: threshold, $t(i) = \frac{G_i * (i * \Delta)}{\max\{G\}}$.

The above algorithm will ensure that the threshold set takes the form of Eq. (21), and will also ensure that the set of thresholds is within the range of the initial unstructured channel estimate as described in [11]. What is lucrative is that $G$ can be a simple look up table if $M$ and $N$ are known; and with an initial estimate, a near optimum set of thresholds can be determined.

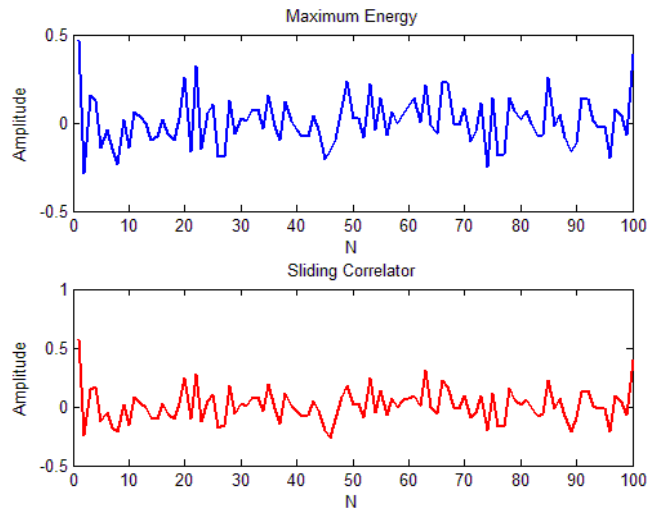

Figure 4. Initial estimates of the channel using two different methods

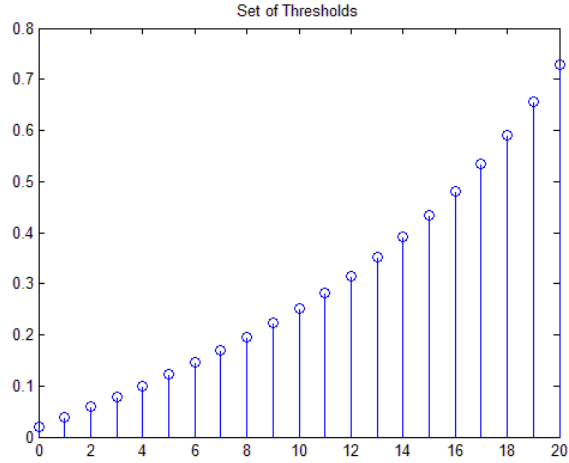

Figure 5. The set of thresholds to be used for estimation using HN algorithm.

For $M = 200$, $N = 100$, $S = 3$, $K = N/2 = 50$, and an initial estimate of the form shown in Fig. 4, $t(i)$ was obtained as shown in Fig. 5. The SNR range of interest was from 0 to 20 $dB$ and for ease of calculation, $N_t = 21$.

### 4. Estimation Process

#### A. Parallel Estimation

A simple Parallel Estimation Algorithm using Compressive Sampling (PEA-CS) can now be described for the estimation process using Eq. (6) and the thresholding technique described in Section 3:

- Determine the vector of thresholds $t$ using the method described in Section 3.
- Use these set of thresholds to determine $N_t$ parallel estimates, $\hat{h}_i$, with $i = 1,2,3…,N_t$ using the corresponding $t(i)$ in the algorithm as described by Eq. (6), instead of $G$.
- Using this estimate find the set of errors: $e(i) = |sum(y - A * \hat{h}_i)|$
- Since the vector $A$ is a product of a random measurement matrix and the convolution matrix the above error is not the least for the best estimate (unlike the algorithm in Ref. [11]). However, we can exploit the fact that many estimates are carried out, and the best estimates are close to each other. Hence:

Find $e' = \dfrac{de(i)}{di}$ and choose $\min_{t(i)}[e'(i)]$; $e' \neq 0$ as the threshold which gives the best estimate

As described above, the results obtained in Fig. 2(b) with HN are by *oracle estimation*. That is, the value of $P$ is known before hand as per Fig. 2(a) and then applied to obtain the best estimate. However with PEA-CS, the estimates were found to be exactly the same with the oracle estimator. This can be shown by choosing a random experiment and looking at the thresholds generated by the oracle estimator, and comparing them with those obtained by the process in Section 3. The results are shown in Fig. 6(a). Notice, that almost all the thresholds generated by the oracle estimator (sorted in plot) lie close to those generated by algorithm used in PEA-CS (bold lines in Fig. 6(a) show the matching). This proves that PEA-CS indeed gives the same MSE as the oracle HN estimator.

Figure 6(b), shows comparison of PEA-CS with Iterative Detection/Estimation (IDE) [11] and Matching Pursuit (MP) [9], and the Cramer-Rao Bounds. As mentioned before, the MSE for PEA-CS were the same as the ones obtained with the oracle HN estimator. Note that MP is the best estimator at low SNR; however, as SNR increases, IDE and PEA-CS outperform MP. An added advantage in PEA-CS is that the computation complexity is much lesser than that of IDE ($O(N^2KN_t)$ vs. $O(N^3N_t)$). As $K$ is usually much smaller than $N$, PEA-CS can be preferred over IDE at SNRs 10-20 $dB$. The reason why MP performs well at low SNR is that the number of taps or sparsity $S$ is assumed available, which may not be the case always. IDE and PEA-CS do not assume this, and hence are highly affected by noise at low SNR.

Note, that IDE involves complex computations such as pseudo-inverses and alternate hard and soft estimates (see Ref. [11] for details), which is eliminated in the case of PEA-CS; thus showing its clear advantage over other algorithms. However, PEA-CS performs very poorly at low SNR. This is due to two reasons: 1) the power constraints are harmed in the HN algorithm; and 2) the use of random projection to obtain a signal of lesser rank in the presence of high noise makes the initial estimate completely noisy.

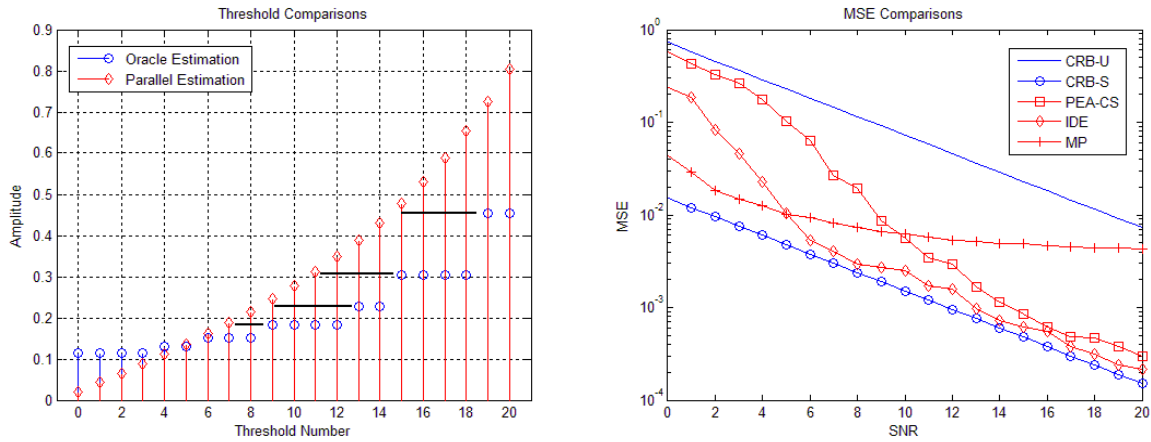

Figure 6. (a) Threshold comparison of the Oracle Estimator and Parallel Estimator (PEA-CS). (b) MSE of PEA-CS compared to IDE, MP and the Cramer Rao Bounds.

## 5. References

[1] Haupt, J. and Nowak, R. (2006). "Signal Reconstruction from Noisy Random Projections", *IEEE Transactions on Information Theory*, September 2006, Vol. 52, No. 9, Pages 4036-4049
[2] Candes, E., and Tao, T. (2005). "The Dantzig Selector: statistical estimation when p is much larger than n", *Annals of Statistics*, 2005, Vol. 3, No. 6, Pages 2313-2351
[3] Candes, E., Romberg, J. and Tao, T. (2006). "Robust uncertainty principles: Exact signal reconstruction from highly incomplete frequency information", *IEEE Transactions on Information Theory*, February 2006, Vol. 52, No. 2, Pages 489-509
[4] Donoho, D.L. (2006). "Compressed Sensing", *IEEE Transactions on Information Theory*, April 2006, Vol. 52, No. 4, Pages 1289-1306
[5] Mallat, S. and Zhang, Z. (1993). "Matching Pursuit with Time-Frequency Dictionaries", *IEEE Transactions on Signal Processing*, December 1993, Vol. 41, No. 12, Pages 3397-3415
[6] Boyd, S. and Vandenberghe, L. (2004). "Convex Optimization", *Cambridge Press*
[7] Candes, E. and Romberg, J. (2005). "l1 magic notes – The Dantzig Selector Implementation", http://www.acm.caltech.edu/l1magic/index.html
[8] Bajwa, U.W., Haupt, J., Raz, G. and Nowak, R. (2008). "Compressed Channel Sensing", *Proceedings of the Annual Conference on Information Sciences and Systems*, March 2008, Pages 1-6
[9] Cotter, S.F. and Rao, B.D. (2002). "Sparse channel estimation via Matching Pursuit with Application to Equalization", *IEEE Transactions on Communications*, March 2002, Vol. 50, No. 3
[10] Karabulut, G.Z. and Yongacoglu, A. (2004). "Sparse channel estimation using orthogonal matching pursuit algorithm". *IEEE $60^{th}$ Conference on Vehicular Technology,* September 2004, Vol. 6, Pages 3880 – 3884
[11] Carbonelli, C. and Mitra, U. (2007). "A Simple Sparse Channel Estimator for Underwater Acoustic Sensors", *IEEE Oceans 2007*, September 2007, Pages 1-6
[12] Candes, E. and Wakin, M.B. (2008). "An Introduction to Compressive Sampling", *IEEE Signal Processing Magazine*, March 2008, Vol. 25, Issue 2, Pages 21-30
[13] Tibshirani, R. (1996). "Regression and selection via the lasso", *Journal of the Royal Statistical Society, Series B*, 1996
[14] Hoeffding, W. (1963). "Probability inequalities for sums of bounded random variables", *Journal of the American Statistical Association*, March 1963, Vol. 58, Pages 13–30
[15] Haupt, J., Bajwa, W., Raz, G. and Nowak, R. "Toeplitz Compressed Sensing Matrices with Applications to Sparse Channel Estimation" *to be submitted*
[16] Edelman, A. (1989). "Eigenvalues and Condition Numbers of Random Matrices", *Thesis, Department of Mathematics, Massachusetts Institute of Technology*, May 1989
[17] Stajanovic, M. (2005). "Retrofocusing techniques for high rate acoustic communications", *Journal of Acoustic Society of America*, March 2005, Vol. 117(3), Part 1, Pages 1173-1185